\pdfoutput=1
\newif\ifFull
\Fullfalse
\newif\ifAnon
\Anonfalse
\ifFull
\documentclass[11pt]{article}
\else
\documentclass[twocolumn]{article}
\fi
\advance \topmargin by -\headheight
\advance \topmargin by -\headsep
\evensidemargin \oddsidemargin
\usepackage{color}
\usepackage{graphicx}
\usepackage{url}
\usepackage{epstopdf}
\usepackage{subfigure}
\usepackage{verbatim}
\usepackage{array}

\makeatletter
\def\@begintheorem#1#2{\sl \trivlist \item[\hskip \labelsep{\bf #1\ #2:}]}
\def\@opargbegintheorem#1#2#3{\sl \trivlist
      \item[\hskip \labelsep{\bf #1\ #2\ #3:}]}
\makeatother

\begin{document}

\title{Privacy-Preserving Group Data Access \\ 
       via Stateless Oblivious RAM Simulation} 

\ifAnon
\author{
Anonymous submission to IEEE Security \& Privacy}
\else

\author{
{Michael T. Goodrich} \\
Dept.~of Computer Science \\ 
University of California, Irvine \\
\vspace{0.1in}
\url{http://www.ics.uci.edu/~goodrich/} \\
Michael Mitzenmacher \\
Dept.~of Computer Science \\ 
Harvard University \\
\url{http://www.eecs.harvard.edu/~michaelm/} \\
\and
Olga Ohrimenko \\
Dept.~of Computer Science \\
Brown University \\
\vspace{0.1in}
\url{http://www.cs.brown.edu/~olya/} \\
Roberto Tamassia \\
Dept.~of Computer Science \\
Brown University \\
\url{http://www.cs.brown.edu/~rt/}
}
\fi

\date{}

\maketitle 


\begin{abstract}
We study the problem of providing privacy-preserving access to an
outsourced honest-but-curious data repository for a group of 
trusted users. 
We show that such privacy-preserving data access is possible using a
combination of probabilistic encryption, which directly hides data values, and
stateless oblivious RAM simulation, which hides the pattern of data
accesses.
We give simulations that have only an $O(\log n)$ amortized time
overhead for simulating a RAM algorithm, $\cal A$, that has a memory of
size $n$, using a scheme that is data-oblivious
with very high probability assuming the simulation has access to a
private workspace of size $O(n^\nu)$, for any given fixed constant~$\nu>0$.
This simulation makes use of pseudorandom hash functions and is based on a
novel hierarchy of cuckoo hash tables that all share a common stash.
We also provide results from
an experimental simulation of this scheme, showing its practicality.
In addition, in a result that may be of some theoretical interest,
we also show that one can eliminate the dependence on pseudorandom
hash functions in our simulation while having the overhead rise to 
be $O(\log^2 n)$.
\end{abstract}



\section{Introduction}
\label{sec:intro}

Companies offering outsourced data storage services 
are defining a growing industry,
with competitors that
include Amazon, Google, 
and Microsoft, which are
providing outsourced data repositories 
for individual or corporate users, with prices that amount to
pennies per gigabyte stored.

Clearly, the customers of such cloud computing services have an
interest in security and privacy, particularly for proprietary data.
As a recognition of this interest,
we note that, as of November 2010, the Amazon S3
and Microsoft Azure cloud platform 
have achieved ISO 27001 certification and Google's cloud computing service
has SAS70 certification.
In spite of these certifications,
the companies that provide outsourced data services nevertheless
often have commercial interests in learning information
about their customers' data.
Thus, the users of such systems
should also consider technological solutions for maintaining the privacy
of their outsourced data in addition to the assurances that come from
certifications and formal audits.

Of course, a key component for
users to maintain the privacy of their data is for them
to store their data in encrypted form, e.g., using
a group key known only to the set of users.
Simply encrypting the group's data is not sufficient to
achieve privacy, however,
since information about the 
data may be leaked by the pattern in which the users access it.
For example, at the Oakland 2010 conference,
Chen {\it et al.}~\cite{cwwz-sclwa-10} show that
highly sensitive data, such as financial and health information,
can be inferred from access patterns at popular
financial and health web sites even if
the contents of those communications are encrypted.

\subsection{Group Access to Outsourced Data}
In this paper, we are interested in technological solutions 
to the problem of protecting the privacy of a group's data 
accesses to an outsourced data storage facility.
In this framework,
we assume that a trusted group, $G$, of users shares a group key, $K$, with
which they encrypt all their shared data that is stored at a
semi-trusted data outsourcer, Bob.
Furthermore,
we assume that the users access their data according to a public
indexing scheme, which Bob knows; hence, we can model Bob's memory, $M$,
as in the standard RAM model (e.g., 
see~\cite{ahu-dsa-83,clrs-ia-01,gt-adfai-02,kt-ad-05}).

Each time a user, Alice, in $G$, accesses Bob's memory, she specifies
an index $i$, and Bob responds with $C=M[i]$.
Alice then performs the following (atomic) sequence of operations:
\begin{enumerate}
\item
She decrypts $C$ using $K$, producing the plaintext
value, $P=D_K(C)$,
that was stored in encrypted form at index $i$ by Bob.
\item
She optionally changes the value of $P$, depending on the
computation she is performing, producing the plaintext value, $P'$.
\item
She encrypts $P'$
using a probabilistic encryption scheme based on $K$, 
producing ciphertext $C' = E_K(P')$.
\item
She returns
$C'$ to Bob for him to store back in his memory 
at index $i$; that is, she directs
Bob to assign $M[i]\leftarrow C'$.
\end{enumerate}
By using a probabilistic encryption scheme, the users in 
the group $G$ ensure
that Bob is computationally unable to 
determine the plaintext of any memory cell from that cell's
contents alone. Also, it is unfeasible for Bob to determine whether
two memory cells store encryptions of the same plaintext.

\subsection{Stateless Oblivious RAM Simulation}
In addition to using probabilistic encryption, 
the users in the group $G$ also need to hide their
data access patterns from Bob, so as to avoid inadvertent information leaks.
To facilitate such information hiding, we formulate the privacy
objective of the users in $G$ in terms
of the \emph{stateless oblivious RAM simulation} problem.

In this framework, we model the group $G$ as a single user, Alice, 
who has a register holding the key $K$ and
a CPU with a private cache.
Alice's interactions with Bob occur in discrete
\emph{episodes} in which she reads
and writes a set of cells in his memory, using probabilistic
encryption, as described above, to hide data contents.
Alice's cache may be used as a private workspace
during any episode, but it cannot store any information from one
episode to the next.
This requirement is meant to model the fact that Alice is
representing a group of users who do not communicate outside of their
shared access to Bob's memory. That is, each episode could model a
consecutive set of accesses from different users in the group $G$.
Moreover, this requirement is what makes this framework
``stateless,''
in that no state can be carried from one episode to the next (other
than the state that is maintained by Bob).

To allow the group of users, which we model by the stateless Alice,
to perform arbitrary computations on the data they share and
outsource to Bob, we assume that Alice is simulating a RAM
computation.
We also assume
the service provider, Bob, is trying to learn as much as possible about the
contents of Alice's data from
the sequence and location of all of Alice's memory accesses. 
As mentioned above, however,
he cannot see the content of what is read or
written (since it is probabilistically encrypted).
Moreover, Bob has no access to Alice's private cache.
Bob is assumed to
be an \emph{honest-but-curious} adversary~\cite{gmw-hpamg-87}, 
in that he correctly
performs all protocols and does not tamper with data.

We say that Alice's sequence of memory accesses is \emph{data-oblivious} 
if the
distribution of this sequence depends only on $n$, the size of the memory
used by the
RAM algorithm she is simulating, $m$, the size of her private
cache, and the length of the access sequence itself. 
In particular, the distribution of Alice's memory accesses
should be independent of the data values in the input.
Put another way, this definition 
means that $\Pr(S\, |\, {M})$,
the probability that Bob sees an access sequence, $S$,
conditioned on a specific configuration of his memory, $M$, satisfies
\[
\Pr(S\, |\, {M}) = \Pr(S\, |\, {M}'),
\]
for any memory configuration
${M}' \not= {M}$ such that $|{M}'|=|{M}|$.

Examples of data-oblivious access sequences for an array, $A$,
of size $n$, in Bob's memory, 
include the following:
\begin{itemize}
\item
Scanning $A$ from beginning to end, accessing each item
exactly once, for instance, to compute the minimum value in $A$,
which is then stored in $A[1]$.
\item
Simulating a Boolean 
circuit, $\cal C$, with its inputs taken in order from the bits of $A$.
\item
Accessing the cells of $A$ according to a random hash function, $h(i)$,
as $A[h(1)]$, $A[h(2)]$, $\ldots$, $A[h(n)]$, or random permutation,
$\pi(i)$,
as $A[\pi(1)]$, $A[\pi(2)]$, $\ldots$, $A[\pi(n)]$.
\end{itemize}
Examples of computations on $A$ that would \emph{not} be data-oblivious 
include the following:
\begin{itemize}
\item
Scanning $A$ from beginning to end, accessing each item
exactly once, to compute the index $i$ of the minimum value in $A$,
and then reading $A[i]$ and writing it to $A[1]$.
\item
Using a standard heap-sort, merge-sort, or quick-sort algorithm to sort $A$.
(None of these well-known algorithms is data-oblivious.)
\item
Using values in $A$ as indices for a hash table, $T$,
and accessing them
as $T[h(A[1])]$, $T[h(A[2])]$, $\ldots$, $T[h(A[n])]$, where $h$ is a
random hash function.
For example, consider what happens if the values in $A$ are
all equal and how unlikely the resulting collision
in $T$ would be.
\end{itemize}
Note that
this last example access pattern 
actually 
would be data-oblivious if the elements in $A$ 
were always guaranteed to be distinct,
assuming the random hash function, $h$, satisfies
the standard assumptions of the random oracle model 
(e.g., see~\cite{br-roap-93}).

\subsection{Related Prior Results}
Data-oblivious sorting is a fundamental 
problem (e.g., see Knuth~\cite{k-ss-73}), with
deterministic schemes giving rise to sorting networks, such as the
impractical
$O(n\log n)$ AKS network~\cite{aks-osn-83,aks-scps-83,p-isnod-90,s-snlgf-09}
as well as practical, but theoretically-suboptimal,
sorting networks~\cite{l-ipaaa-92,p-ssn-72}.
Randomized data-oblivious sorting algorithms running in
$O(n\log n)$  time and succeeding with high probability%
\footnote{In this paper, we take
  the phrase ``with very high probability'' to mean that the
  probability is at least $1-O(1/n^d)$, for any given fixed constant $d\ge 1$.}
are
studied by Leighton and
Plaxton~\cite{lp-hsn-98} and Goodrich~\cite{g-rsaso-10}.
In addition, data-oblivious sorting is finding applications to
privacy-preserving secure multi-party
computations~\cite{wlgdz-bpstp-10}, and it is used in all the
known oblivious
RAM simulation schemes (including the ones in this paper).

In early work on the topic of oblivious simulation,
Pippenger and Fischer~\cite{pf-racm-79} show that one can simulate a
Turing machine computation of length $n$ with
an oblivious Turing machine computation of length $O(n\log n)$, that
is, they achieve an amortized $O(\log n)$ time and space
overhead for this oblivious simulation.

\begin{table*}[ht]
\begin{center}
\caption{Comparison of Oblivious RAM simulations.}
\begin{tabular}{|c|>{\centering}m{1.7cm}|>{\centering}m{1.89cm}|c|>{\centering}m{3cm}|}
\hline
& User Memory & User State Size  & Server Storage & Amortized Access Overhead
\tabularnewline
\hline
Goldreich and Ostrovsky~\cite{go-spsor-96}  &$O(1)$& - & $O(n\log n)$& $O(\log^3n)$
\tabularnewline
Williams and Sion~\cite{DBLP:conf/ndss/WilliamsS08} & $O(\sqrt{n})$ & $O(\sqrt{n})$ & $O(n\log n)$ & $O(\log^2n)$
\tabularnewline
Goodrich and Mitzenmacher~\cite{gm-paodor-11} (1) & $O(1)$ &  - & $O(n)$& $O(\log^2n)$
\tabularnewline
Goodrich and Mitzenmacher~\cite{gm-paodor-11} (2) & $O(n^\nu)$ &  $O(n^\nu)$ & $O(n)$& $O(\log n)$
\tabularnewline
Boneh {\it et al.}~\cite{bmp-rosmor-11} & $O(\sqrt{n\log n})$ &  $O(\sqrt{n\log n})$ & $O(n)$& $O(1)$
\tabularnewline
\hline
\hline
Our result &  $O(n^\nu)$ & - &  $O(n)$& $O(\log n)$
\tabularnewline
Our result (w/o random oracle) &  $O(n^\nu)$ & - &  $O(n)$&$O(\log^2 n)$
\tabularnewline
\hline
\end{tabular}
\end{center}
\label{tbl:summary}
\end{table*}

More recently,
Goldreich and Ostrovsky~\cite{go-spsor-96} show that a
RAM computation using space $n$ can be simulated 
with an oblivious RAM with an amortized time overhead of $O(\log^3 n)$ 
per step of the original RAM algorithm and space
overhead of $O(\log n)$.
Goodrich and Mitzenmacher~\cite{gm-paodor-11} improve this result
by showing that any
RAM algorithm, $\cal A$, can be simulated in a data-oblivious 
fashion, with very high probability,
in an outsourced memory
so that each 
memory access performed by $\cal A$ has a time overhead of
$O(\log^2 n)$, assuming Alice's private cache has size $O(1)$.
Their scheme has a space overhead of $O(1)$.
Incidentally,
in the recent CRYPTO 2010 conference,
Pinkas and Reinman~\cite{pr-orr-20}
also claim an oblivious RAM simulation
result having a time overhead of $O(\log^2 n)$, but there is a flaw in
this version of their 
scheme\footnote{The scheme of Pinkas and Reinman allows the
adversary, Bob, to distinguish with high probability an access sequence
that reads the same memory location over and over from one that
accesses each memory cell exactly once. This flaw is expected to 
be repaired in the journal version of their paper.}.

In addition to these stateless oblivious RAM simulation schemes,
Williams and Sion~\cite{DBLP:conf/ndss/WilliamsS08}
show how to simulate a RAM computation with an
oblivious RAM where the data
owner, Alice, has a
stateful private memory of size $O(\sqrt{n})$,
achieving an expected amortized time overhead of $O(\log^2 n)$ using
$O(n\log n)$ memory at the data provider.
In addition,
Williams {\it et al.}~\cite{wsc-bcomp-08} claim a
  method that uses an $O(\sqrt{n})$-sized private cache
  and has $O(\log n\log\log n)$ amortized time overhead,
but Pinkas and Reinman~\cite{pr-orr-20}
  have raised concerns with the assumptions and analysis of
  this result.

Goodrich and Mitzenmacher~\cite{gm-paodor-11} provide a stateful 
RAM simulation scheme that
achieves an overhead of $O(\log n)$ and is oblivious with very high
probability.
Their scheme assumes that Alice maintains state from one episode to
the next in a private cache of size $O(n^{\nu})$, for any given fixed constant
$\nu>0$.
Boneh {\it et al.}~\cite{bmp-rosmor-11} also propose a scheme
that uses a state. They achieve an amortized overhead
of $O(1)$ but using a state of size $O(\sqrt{n\log n})$.
However, this state is essential to the efficiency of both
simulation schemes.
Thus, these methods are not applicable to the problem of providing
privacy-preserving group 
access to an outsourced data repository.

Returning to stateless oblivious RAM simulation, we note that
Ajtai~\cite{a-orwca-10} has a recent oblivious RAM simulation result
that shows that a polylogarithmic
factor overhead in time and space is possible
without cryptographic assumptions about the existence of
random hash functions, as is done in the previous oblivious RAM
simulation cited above.
Damg\aa{}rd {\it et al.}~\cite{dmn-psor-10} improve this result
further, 
showing that a time overhead of $O(\log^3 n)$ is possible for
oblivious RAM simulation without using random functions.

In addition to the above-mentioned upper-bound results,
Beame and Machmouchi~\cite{bm-mrors-10} show that if the
additional space utilized in the simulation (besides the space
for the data itself) is sufficiently sublinear, then the overhead for
oblivious RAM simulation has a superlogarithmic lower bound.
Such bounds don't apply, of course, to a simulation that uses $O(n)$
additional memory, as is common in the efficient schemes mentioned
above.

We provide a summary of the Oblivious RAM simulation schemes and
compare with ours in Table~\ref{tbl:summary}. Note that the schemes that
maintain a state cannot be used to hide a pattern of access by a group of
users which is one of the challenges we address in this paper.

\subsection{Our Results}
We give an efficient method for simulating
any RAM algorithm, $\cal A$, in a stateless fashion
with a time overhead of $O(\log n)$ 
and space overhead of $O(1)$, using an access sequence that is
data-oblivious with very high probability, where $n$ is the size of the RAM memory.
Our methods assume that Alice has a private cache of of size
$O(n^{\nu})$, for any given fixed constant $\nu>0$, but she uses this cache only as a
private ``scratch space'' to support computations she performs during
each episode.
Alice is not allowed to maintain state in her private memory from one
episode to the next.
Thus, this simulation scheme is applicable to the problem of
simulating access to a shared data repository by a group of
cooperating users that all share a secret key.
Moreover, the assumption about the size of Alice's scratch space is
motivated by the fact that even handheld devices have a reasonable
amount of local memory.
For example, if we were to set $\nu=1/4$, 
then our simulation would allow a collection
of devices having memories with sizes on the 
order of one megabyte to support privacy-preserving access to
an outsourced data repository whose size 
is on the order of one yottabyte.

Like the previous oblivious RAM simulation schemes 
mentioned above, our
scheme uses a hierarchy of hash tables, together with a small set of
pseudorandom hash functions, to obfuscate the access
pattern of the algorithm $\cal A$ (which need not be specified in
advance).
The main idea of our scheme is to maintain these hash tables
as cuckoo hash tables that all share a single stash of size $O(\log n)$.
While conceptually simple, this approach requires a new, non-trivial
analysis for a set of cuckoo tables sharing a common stash.
In addition, an important technical detail that
simplifies our construction is that we make no use of so-called
``dummy'' elements, whereas the previous schemes used such elements.

In practice, the set of pseudorandom hash functions could be implemented
using, e.g., keyed SHA-256 functions~\cite{dp-gboehf-08}.
Nevertheless,
we also show that our construction can be used to simulate a RAM
computation with an overhead of $O(\log^2 n)$ without the use of
pseudorandom functions, which may be of some theoretical interest.

Finally, we provide experimental results for a simulation of our
scheme, which show the practical effectiveness of the approach of using a
shared stash.
In particular,
our experimental prototype simulates the dynamic evolution of the
hierarchy of hash tables and
our experimental analysis shows the 
threshold values at which the shared stash becomes effective.


\section{Theory Background}
\label{sec:background}

For our results, we rely on general methods for data-oblivious
simulation of a non-oblivious algorithm on a RAM.  
As mentioned above, the seminal
theoretical framework for such simulations was presented by Goldreich
and Ostrovsky~\cite{go-spsor-96}, who store keys in a hierarchy of
hash tables of increasing size, each being twice the size of the
previous one.  For $n$ items there are $O(\log n)$ levels, each level
being a standard hash table with $2^i$ buckets for some $i$, and each
bucket containing up to $O(\log n)$ keys in order to cope with
collisions within the hash table.  In this construction the total size
of all the tables is $O(n \log n)$.  To perform a lookup, the first level
is scanned sequentially, and in each of the other levels, a bucket
chosen by the hash function for that level acting on the key (or, if
the item is found at an earlier level, a random dummy key) is scanned.
The item is subsequently re-encrypted and re-inserted into the first
level.  It is important to note that at all levels a bucket is scanned
even if the key is found early, to maintain obliviousness.  As levels
fill, keys must be shifted down to subsequent levels.  The details of
the original scheme are rather complex; for further details see the
original paper~\cite{go-spsor-96}.

Recently, a more efficient simulation approach for this problem was
outlined by Goodrich and Mitzenmacher~\cite{gm-paodor-11}.  The primary
difference in this new line of work is the use of {\em cuckoo hash
  tables} in place the standard hash tables used originally in
\cite{go-spsor-96}.  We therefore now present some background on {\em
  cuckoo hashing}.

As introduced by Pagh and Rodler \cite{pr-ch-04}, in standard cuckoo
hashing we utilize two tables, each with $m$ cells, with each cell
capable of holding a single key.  We make use of two hash functions
$h_1$ and $h_2$.  We assume that the hash functions can be modeled as
completely random hash functions.  The tables store up to $n$ items,
where $m = n (1+\epsilon)$ for some constant $\epsilon > 0$, yielding a load of
(just) less than $1/2$; keys can be inserted or deleted over time as
long as this restriction is maintained.

\begin{figure}[ht]
  \begin{center}
    \includegraphics[scale=0.30]{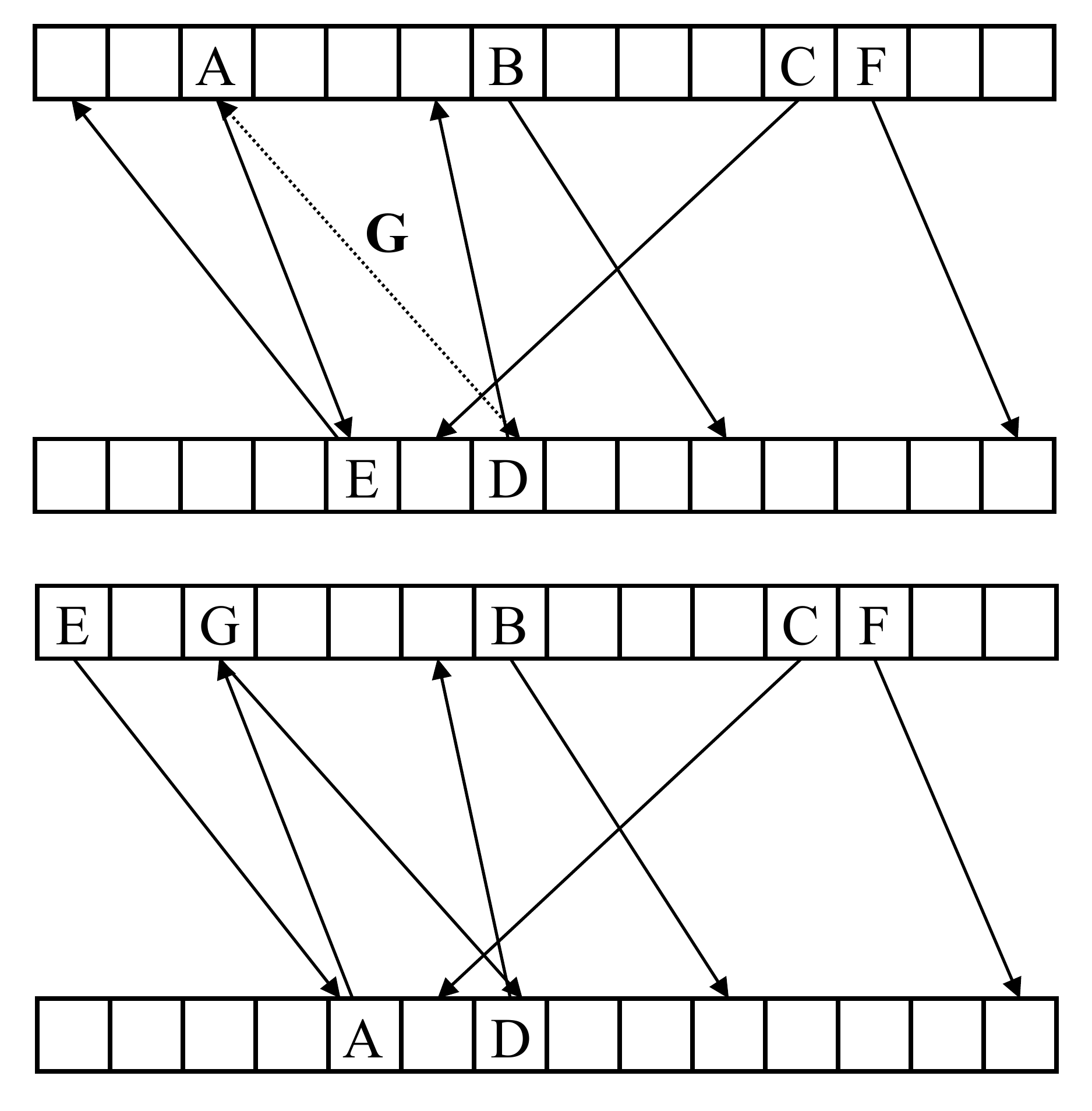}
  \end{center}
  \caption{The top of the figure represents a cuckoo hash table.  Keys are placed in one subtable;  the arrow
for each key points to the alternate location for the key in the other subtable.  Key G is inserted, leading to the movement of several other keys for G to be placed, as shown in the bottom of the figure.}
  \label{fig:cuckoo1}
\end{figure}

\begin{figure}[htp]
  \begin{center}
    \includegraphics[scale=0.30]{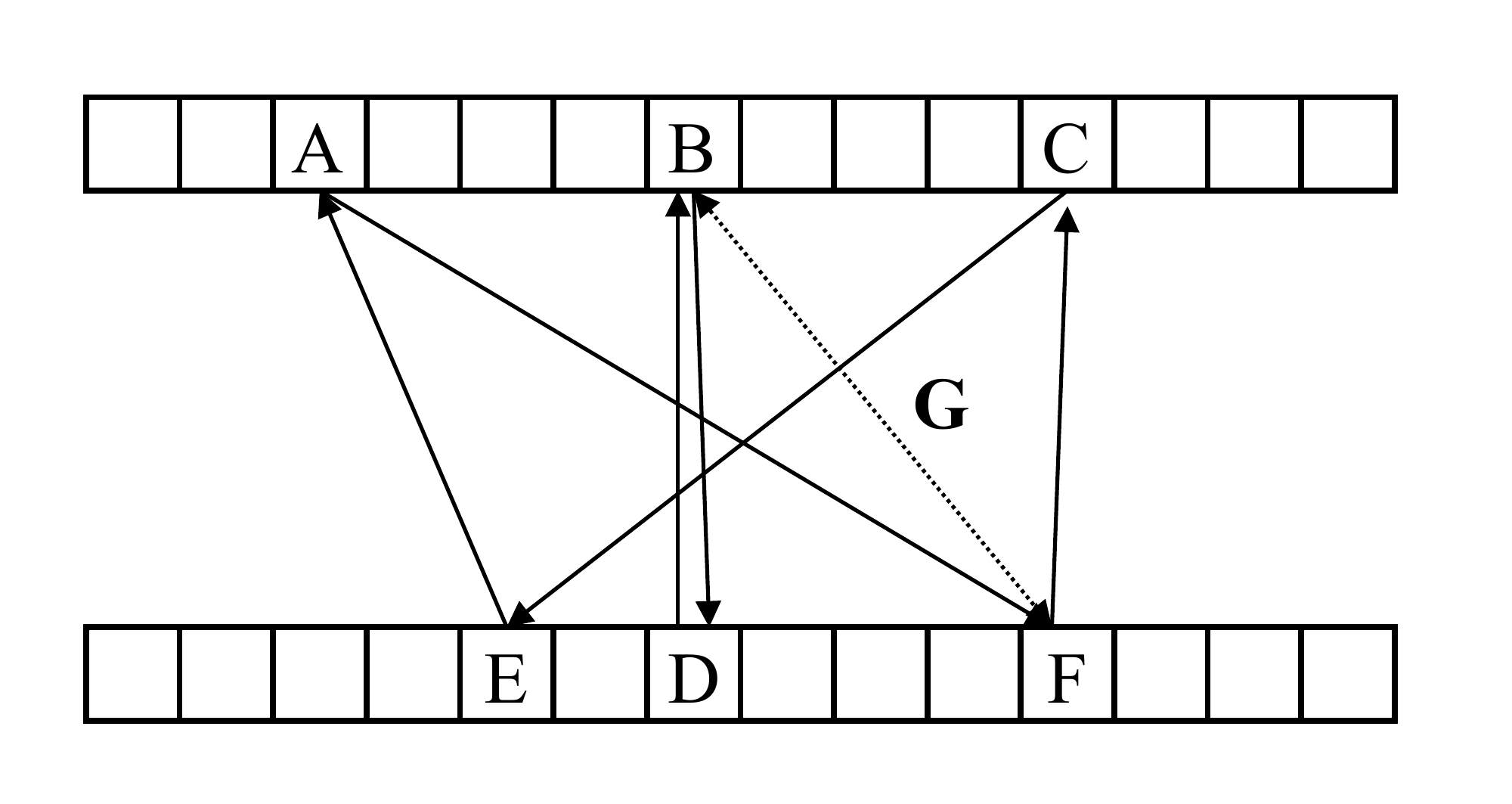}
  \end{center}
  \caption{Key G is to be inserted, but it cannot be placed successfully.  (Seven keys have only six locations.)
This leads to a failure, or if there is a stash, then G can be placed in a stash.}
  \label{fig:cuckoo2}
\end{figure}

A key $x$ (which we may also refer to as an ``item'' or ``element'')
that is stored in the hash tables must be located at either $h_1(x)$
or $h_2(x)$.  As there are only two possible locations for a key,
lookups take constant time.  To insert a new key $x$, we place $x$ in
the cell $h_1(x)$.  If the cell had been empty, the operation is
complete.  Otherwise, key $y$ previously in the cell is moved to
$h_2(y)$.  This may in turn require another key to be moved, and so
on, until a key is placed in an empty cell.  We say that a failure
occurs if, for an appropriate constant $c$, after $c \log n$ steps
this process has not successfully terminated.  Suppose we insert an $n$th key into the system.
It is known that:
\begin{itemize}
\item The expected time to insert a new key is bounded above by a constant (that depends on $\epsilon$).
\item The probability that a new key causes a failure is $\Theta(1/n^2)$ (where the notation hides
a dependence on $\epsilon$).  
\end{itemize}
See Figures~\ref{fig:cuckoo1} and~\ref{fig:cuckoo2} for examples.  

Before considering ways to reduce the probability of failures to
something more suitable, we briefly mention that there are several
natural variations of cuckoo hashing, many of which are described in
a survey article by Mitzenmacher~\cite{m-soq-09}.  Variations include using more
than two choices, using cells that hold more than one key, and so on.
For our purposes, it suffices to understand standard cuckoo hashing,
along with idea of a {\em stash} \cite{kmw-chs-09}.  

A stash
represents additional memory where keys that would cause a failure can
be placed in order to avoid the failure; with a stash, a failure
occurs only if the stash itself overflows.  As shown in
\cite{kmw-chs-09}, the failure probability when inserting the $n$th
key into a cuckoo hash table can be reduced to $O(1/n^{k+2})$ for any constant $k$ by using a
stash that can hold $k$ keys.  Using this allows us to use cuckoo hash tables
for any polynomially bounded number of inserts and deletions using only
a constant-sized stash.  To search for an item, we must search both the two table
locations and the $k$ stash locations.  In the context of oblivious simulation,
we can search the stash simply by reading each stash location.  

As we have stated, however, in order to perform our oblivious
simulation, we will make use of a hierarchy of cuckoo hash tables to
hold $n$ items.  The smallest of these hash tables may be much smaller
than $n$, which can lead to problems in our setting\footnote{This is
  also at the heart of the flaw in the CRYPTO2010 version of the
  Pinkas and Reinman paper~\cite{pr-orr-20}.}.  
For example, if
the smallest hash table is of size $x$, then even using a stash of size $k$
leads to a failure probability of $O(1/x^{k+2})$.  If $x$ is for
example polylogarithmic in $n$, then for any constant $k$, the failure
probability is $\Omega(1/n)$, and therefore over the insertion of $n$
items, we would expect failures to occur.  In order to deal with this
problem, Goodrich
and Mitzenmacher~\cite{gm-paodor-11} extend the analysis of \cite{kmw-chs-09}
to stashes of logarithmic size, showing that even for suitably large table sizes $x$
that are only polylogarithmic in $n$, and stashes of size $k = O(\log n)$, the
failure probability is $O(x^{-\alpha k})$ for a suitable constant $k$.  This 
suffices to yield superpolynomially small failure rates.  

In fact, we need to extend this result even further here.  In
\cite{gm-paodor-11}, Goodrich and Mitzenmacher use a logarithmic-sized
stash at each level.  We explain here that it suffices to use a
{\em single} logarithmic-sized stash for all levels.  That is, while
there's a non-trivial probability of {\em at least one} layer in our
hierarchy requiring a stash of size $\Omega(1)$, over the logarithmic
number of layers only a stash of logarithmic size is actually
necessary.  We will use this in our construction in Section~\ref{sec:oram}.
We now briefly explain why, if we consider the sum of
the number of items placed in the stash at all possible levels in our
construction, this will be at most $O(\log n)$ with high probability.  

The key is the following argument.  As shown
in \cite{gm-paodor-11}, at a level of size $x$ cells (where $x$
is $\Omega(\log^7 n)$), the probability that the stash for that level
exceeds a total size $s$ is $x^{-\Omega(s)}$.  Further, as long as the
hashes for each level in our construction are independent, we can
treat the required stash size at each level is independent, since the
number of items placed in the stash at a level is then a random
variable dependent only on the number of items appearing in that
level.

Now consider any point of our construction and let $S_i$ be the
number of items at the $i$th level that need to be put in the stash.
It is apparent that $S_i$ has mean less than 1 and tails that can be
dominated by a geometrically decreasing random variable.  This is
sufficient to apply standard Chernoff bounds.  Formally, let $X_1,
X_2, \ldots, X_\ell$ be independent random variables with mean 1
geometrically decreasing tails, so that $X_i = j$ with probability
$1/2^{j}$ for $j \geq 1$.  Then the calculations
of \cite{gm-paodor-11} imply that the $X_i$ stochastically
dominate the $S_i$, and we can now apply standard Chernoff bounds for
these random variables.  Specifically, noting that $X_i$ can be
interpreted as the number of fair coin flips until the first heads, we
can think of the sum of the $X_i$ as being the number of coin flips
until the $\ell$th head, and this dominates the number of items that
need to be placed in the stash at any point.  When $\ell = O(\log n)$,
as is the case here as there are only $O(\log n)$ levels of hash
tables in our construction, then for any constant $\gamma_1$ there
exists a corresponding constant $\gamma_2$ such that the $\ell$th head
occurs by the $(\gamma_2 \log n)$'th flip with probability at least
$1-1/n^{\gamma_1}$.  (See, for example, \cite[Chapter 4]{mu-pcrap-05}.) 
Hence we can handle any polynomial number of steps
with high probability, using a stash of size only $O(\log n)$ that
holds items from all levels of our construction.

\section{Simulating a RAM Algorithm Obliviously}
\label{sec:oram}
In this section, we describe and analyze two schemes for stateless
oblivious RAM simulation.

\subsection{Simulation Using Pseudorandom Functions}
\label{sec:simulator1}

We begin with a construction that uses pseudorandom functions and is secure
against a polynomially bounded adversary.
 
Given a RAM
algorithm, $\cal A$, the main goal of our oblivious simulation of $\cal A$
is to hide the pattern of memory accesses that are made by~$\cal A$.  As
mentioned in Section~\ref{sec:background}, we follow the general
framework introduced by Goldreich and Ostrovsky~\cite{go-spsor-96}, which
uses a hierarchy of hash tables.
 
\begin{figure*}[hbt!]
  \centering
  \includegraphics[scale=0.5]{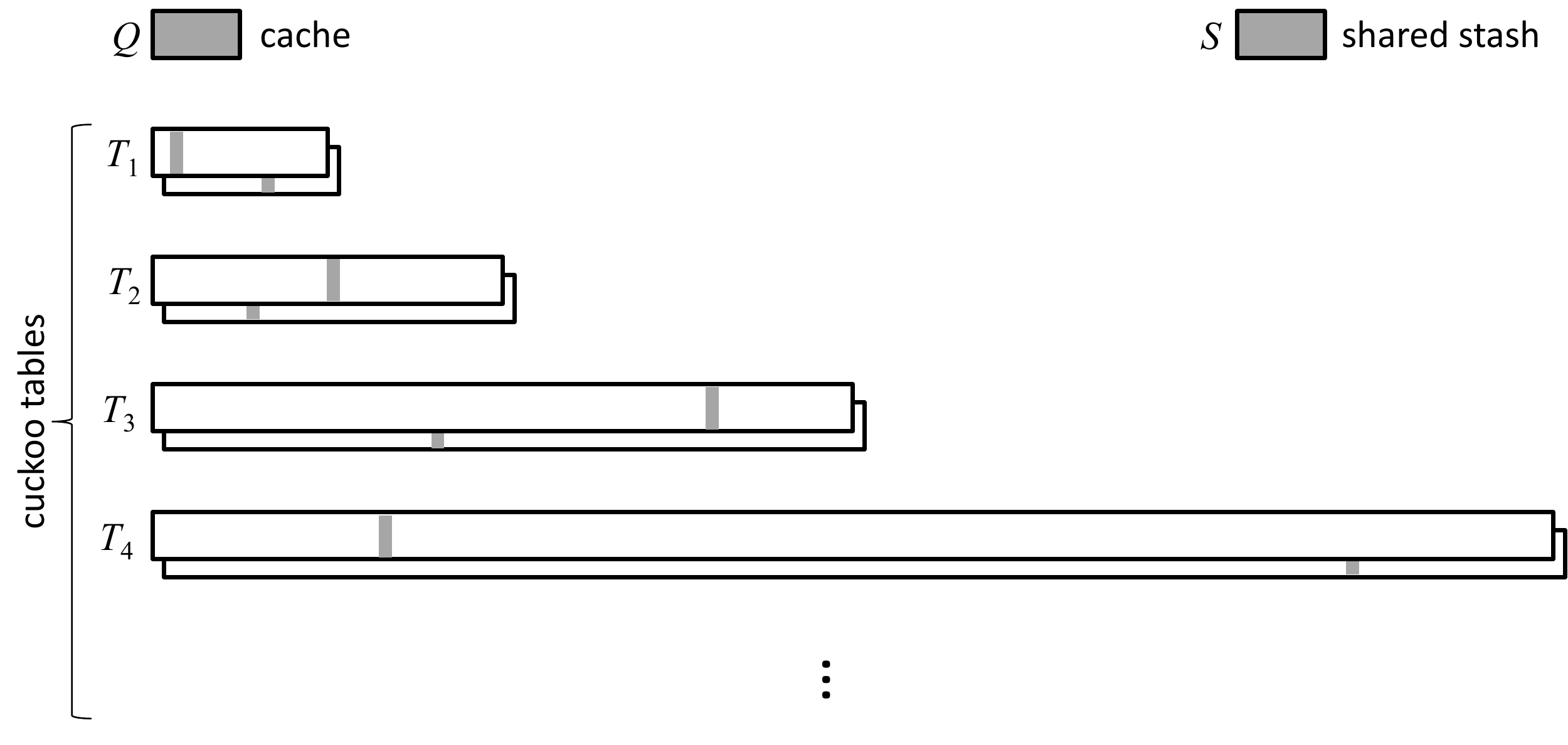}
  \caption{ \label{fig:hierarchy}%
    Illustration of the data structure stored at the server for oblivious
    RAM simulation using pseudorandom functions. In the access phase of the
    simulation, all the items in the cache, $Q$ and the stash, $S$, plus
    two items for each cuckoo table $T_i$ are read by the server. The
    locations accessed by the server are visualized as gray-filled
    rectangles.  }
\end{figure*}

Let $n$ be the number of memory cells of the RAM. 
We view each such cell as an
item consisting of a pair
$(x,v)$, where $x \in \{0, \cdots, n-1\}$ is the index and $v$ is the
corresponding value. Our data structure stored at the server has three
components, illustrated in Figure~\ref{fig:hierarchy}.  The first component is a cache of size $O(\log n)$, denoted by~$Q$.
The second component is a hierarchy of cuckoo hash tables,
$T=(T_1,\ldots,T_L)$, where the size of $T_1$ is twice
the size of $Q$, each table $T_{i+1}$ is twice the size of table~$T_i$, 
and $T_L$ is the first table in the sequence of size greater than or equal to~$n$. 
Thus, $L$ is $O(\log n)$.
The third component is a stash, $S$, shared between all the above cuckoo tables.

RAM items are stored in the data structure in encrypted form.  We use a
semantically secure probabilistic encryption scheme, which results in a
different ciphertext for the same item each time it is re-encrypted.  Also,
the server is unable to determine whether two ciphertexts correspond to the
same item.  
The stash $S$ is handled
in a similar manner whenever we search in it for an item.

We use
a family of pseudorandom functions parameterized by a secret value,
$k_i$, for each table, $T_i$, such that no value $k_i$ is
revealed to the server.
In particular, $k_i$ is stored in encrypted form for each table $T_i$,
so that each user can read $k_i$, decrypt it, and then use it to
provide the two hash functions, $h_1$ and
$h_2$, employed by the cuckoo table, $T_i$,
to determine the location of items. 
In particular, a memory item $(x,v)$ is mapped
to locations $h_1(x)$ and
$h_2(x)$ in $T_i$ by the cuckoo scheme (and stored in one of these two locations
or in the common stash, $S$).

The data structure is initialized by storing all the $n$ RAM items into
cuckoo table~$T_L$.  Each memory access defined by algorithm $\cal A$
corresponds to an \emph{episode} in our simulation. An episode consists of
two phases, an \emph{access phase} and a \emph{rebuild phase}.

Suppose algorithm $\cal A$ calls for an access to memory item~$(x,v)$.
The access phase consists of a search for $x$ in the cache, $Q$, then in the stash, $S$, and
continues with a two-cell cuckoo lookup 
in each of $T_1$ to $T_L$ until we find the first item
with index~$x$.
Once we have found this item, we have achieved the goal of our
search, of course. 
Nevertheless, for the sake of obliviousness,
we simulate continuing the search throughout the entire data structure. Namely,
we always traverse completely $Q$ and $S$, and we
perform two-cell cuckoo accesses in tables $T_1$ through~$T_L$. However, 
after the item is found, 
we simply access two distinct, independent uniformly chosen random locations in each
remaining cuckoo table.

Once we have completed the access phase, which takes $O(\log n)$ time, we
then switch to the rebuild phase.  We begin by adding or replacing a copy
of the found item into cache~$Q$, possibly changing its value in the case
of a write operation.  
To assure obliviousness, we exhaustively scan $Q$ in
a sequential manner and re-encrypt and rewrite all its items.  Thus, the
server cannot distinguish which item was accessed and whether it was
modified.

We note briefly that if the item is in the stash, we can obliviously
remove it from the stash when placing it into $Q$, to help make sure
the stash does not overflow.  One natural approach is to have stash
cells have an associated ``clean'' or ``dirty'' bit, which is
encrypted along with the rest of the item.  A clean cell can store
an item;  a dirty cell is currently being utilized.  When an item is found
and replaced into $Q$, we can set the cell to clean in the stash.      

After adding enough items, cache $Q$ will eventually overflow. We remedy
the overflow by moving all the elements of $Q$ to cuckoo table~$T_1$,
including those associated with empty locations. However, in order to
maintain obliviousness, we do not wait for an overflow to occur and instead
perform the move after a number of accesses equal to the size of~$Q$. The
moving down of elements cascades down through the hierarchy of cuckoo
tables at a fixed schedule by periodically moving the elements of $T_{i-1}$
into~$T_{i}$ at the earliest time $T_{i-1}$ could have become full.
Also, suppose that we are going to move elements into table $T_i$
for the second time, then we instead move the elements
into table~$T_{i+1}$.
Moreover, we continue applying this rule for $i=1,2,\ldots$, until we
are copying the elements into a table for the first time or we reach~$T_L$.
Thus, the process of copying elements into a cuckoo table occurs at
deterministic instances, depending only on the place we currently are
at in the access sequence specified by algorithm~$\cal A$.

\begin{figure*}[tb!]
  \centering
  \includegraphics[width=6in, trim = 0.2in 1.1in 0.2in 2.5in, clip]{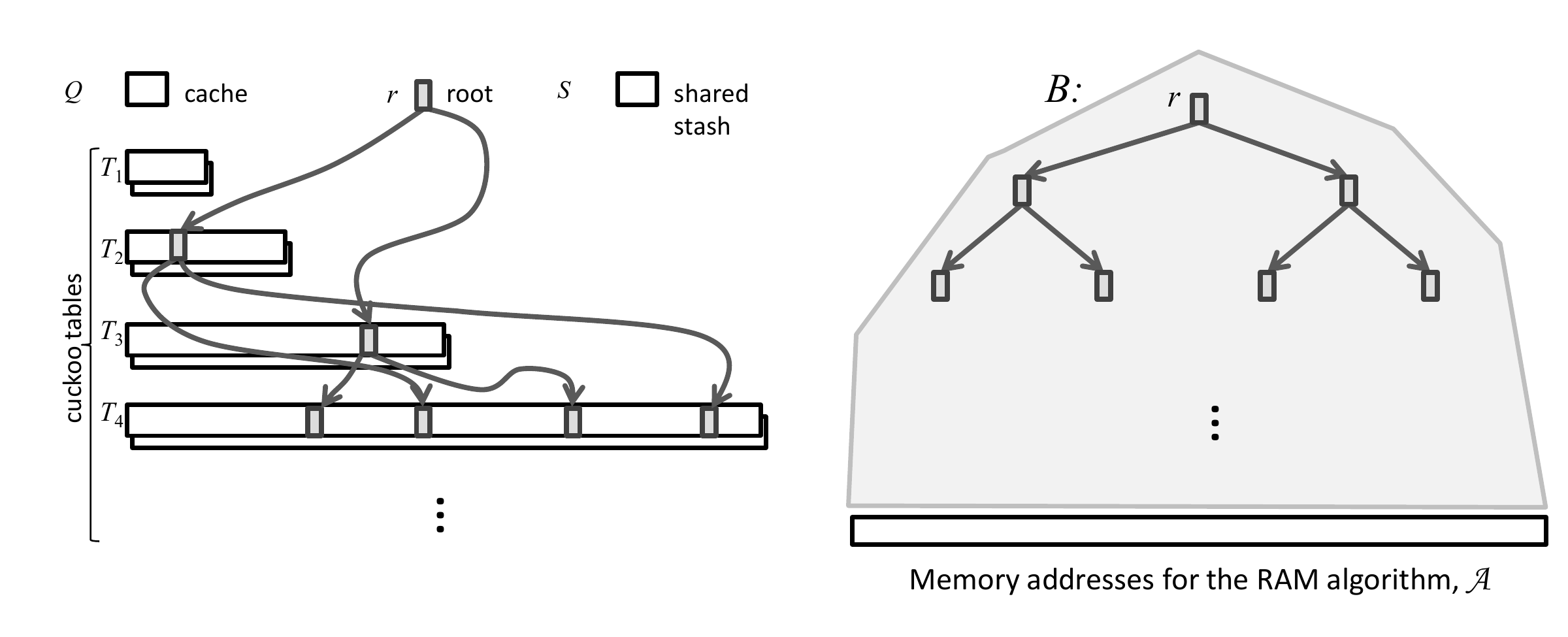}
  \caption{ \label{fig:tree}%
    Illustration of the data structure stored at the server for oblivious
    RAM simulation without using pseudorandom functions. 
    The binary tree is shown conceptually on the right and in terms
    of the storage locations of its nodes on the left.
    The storage locations for the nodes of the binary tree, $B$,
    are visualized as gray-filled
    rectangles.  
    In the access phase of the
    simulation for a non-root node, 
    all the items in the cache, $Q$ and the stash, $S$, plus
    two items for each cuckoo table $T_i$ are read by the server. 
    }
\end{figure*}

In order to move $m$ elements from a table $T_i$
into a cuckoo hash table $T_{i+1}$ obliviously, we
use an algorithm of~\cite{gm-paodor-11}
to obliviously sort the items using $O(m)$ accesses to the outsourced
memory, assuming we have a private workspace of size $O(n^\nu)$,
for some constant $\nu>0$, and $m\ge \log n$, which is always true in
our case.
This allows us to remove duplicate items and
use another algorithm of~\cite{gm-paodor-11}
to obliviously construct a cuckoo table of size $m$ and an 
associated stash, $S'$,
of size $O(\log n)$ in $O(m)$ time, with very high
probability, while utilizing the private workspace of size $O(n^\nu)$.
Given this construction, we then read $S$ and $S'$ into our private
workspace, remove any duplicates and 
merge them into a single stash $S$ (which will succeed with
very high probability, based on the analysis we have given above),
and write $S$ back out in a straightforward oblivious fashion.
Note that in order to assure obliviousness in subsequent lookups, table $T_{i+1}$
is rebuilt using two new pseudorandom hash functions
selected by the client by replacing parameter $k_{i+1}$ with a new
secret value.

Any access performed in our simulation will eventually
lead to $O(\log n)$
table rebuilds, with each element in a rebuild being charged with a
constant amount of work; hence, the total amortized overhead of all rebuild
phases is $O(\log n)$.
Therefore, the total amortized time overhead of the entire simulation
is $O(\log n)$. Moreover, it is easy to see that the space used
by the data structure stored at the server is~$O(n)$.

Let us therefore consider the obliviousness of this scheme.
As we have already observed, the rebuild phase is clearly oblivious,
so any potential dependencies on input values would have to come in
the access phase.
Recall that in the access phase, we search in $S$, $Q$,
and do a two-table cuckoo access in $T_1,\ldots,T_L$.
Moreover, because we move the found item into $Q$ after each
access, and we switch to performing random table lookups once we have
found the item, we are guaranteed never to repeat 
a two-cell cuckoo lookup in any table, $T_i$, for the same item $x$.
In addition, each such lookup is an independent uniformly random
access to a table (either from our assumption about $h_1$ and $h_2$
being distinct random hash functions for each table
or because we already found the item and are
making random accesses explicitly).
We perform $O(|T_i|)$ such lookups before we empty $T_i$; hence, 
the obliviousness of our access sequence depends on the inability of
the adversary, Bob, of telling if we are doing a search for an actual
item or performing a random access for the sake of
obliviousness.
That is, with high probability, Bob should not be able to determine 
whether the item was in $S$, $Q$, or some $T_i$ at the point we found it.
Note that this ability depends solely on whether or not the
$O(|T_i|)$ accesses we made to $T_i$, together with 
searches in the shared stash
$S$, would correspond to valid cuckoo lookups in $T_i$ for some set
of items.
Of course, this is the same as the event that inserting all these
elements into $T_i$ would form a valid cuckoo table, with shared
stash $S$, which we have already observed (in
Section~\ref{sec:background}) is an event that occurs
with very high probability.
Thus, our scheme is oblivious with very high probability.

\subsection{Simulation Without Pseudorandom Functions}
We can adapt our simulation to avoid the use of random functions by
employing an elegant trick due to 
Damg\aa{}rd {\it et al.}~\cite{dmn-psor-10}, 
albeit now further simplified to avoid the use of dummy nodes, which
would add an extra level of complication that our scheme doesn't
require.

The main idea is to place a complete binary tree, $B$, on top of all the
memory cells used in the algorithm $\cal A$, and access 
each memory cell $x$
by performing a binary search from the root of $B$ to the leaf node
corresponding to $x$.
That is, we associate each memory cell item used
by $\cal A$ with a leaf of $B$, define $B$ to have height 
$\lceil\log n\rceil$, and include information at each
internal node $v$ of $B$ so that a search for $x$ 
can determine in $O(1)$ time whether to proceed with the left child
or right child of $v$.
In our case, we store
each of the nodes of $B$ in our hierarchy of tables, 
similar to what is described above, with 
the shared stash, $S$, the cache, $Q$, and the set of cuckoo tables,
$T_1$ to $T_L$.
(See Figure~\ref{fig:tree}.)

The main difference of this scheme with that given above is that in this
case we no longer use random hash functions, $h_1$ and $h_2$, to
determine the locations of each element $x$ in a cuckoo table $T_i$.
Instead, we simply choose two distinct, independent uniformly random
locations, $i_1$ and $i_2$, in the respective two sides of $T_i$ and 
associate these with $x$ as a tuple $(x,i_1,i_2)$, which now
represents the element $x$ in our table.

Initially, all the nodes of $B$ are stored in this way in $T_L$, and
for each such internal node $v$, we include in $v$'s record 
pointers to the two random indices (and table index) for 
$v$'s left child and pointer to the two random indices 
(and table index) for $v$'s right child.
Such pointers can be built obliviously by $O(1)$ calls to oblivious
sorting once we have placed all the nodes into $T_L$.
Moreover, we will maintain such pointers throughout our simulation.
In addition, we store the root $r$ of $B$ separately, as it is
accessed in every step of our simulation.

Let us consider, therefore, how an access now occurs.
The critical property, which we maintain inductively, is that, for
each node $v$ in $B$, which, say, is stored in $T_j$ as its earliest (highest)
location in our hierarchy, all the ancestors of $B$ are stored
in the tables $T_1,\ldots,T_j$, or in $r$, $S$, or $Q$.

Our access for a memory cell $x$ now occurs as a root-to-leaf search
in $B$.
We begin by searching in $r$ to identify the two random indices and
the table index for each of $r$'s children.
Based on the value of $x$, we need to search next for either the left
or right child of $r$, so let $i_1$ and $i_2$ be the two random indices 
for this node, $w$, and let $j$ be the index of the highest
table $T_j$ storing $w$ (with $j=0$ if $w\in Q$ and $j=-1$ if $w\in S$).
We next search in $S$ and $Q$ for $w$, and then proceed 
in $T_1$ through $T_L$.
Of course, we already know the table where we will find $w$.
So, for each table $T_k$ with $k\not= j$,
we simply access two random locations in $T_k$ for the sake of
obliviousness.
For $T_j$ itself, we look in locations 
$T_j[i_1]$
and 
$T_j[i_2]$ to find the cell containing the record for $w$.
If $w$ is not a leaf node, we repeat
the above lookup search for the appropriate child of $w$ that will
lead us to the node storing $x$.

Once we have done our lookup for $x$, and have accessed a
root-to-leaf set of nodes, 
\[
W=\{w_1,w_2,\ldots,w_{\log n}\},
\]
in the process, we perform a rebuild phase for $W$, as
in the above construction based on random hash functions, except that
we use random locations for all the nodes we move rather than use
random functions.
Note that by our induction hypothesis, if we move a set of nodes into
a table $T_i$, then all the pointers for these nodes are either in
$T_i$ itself (hence, can be identified after $O(1)$ calls to
oblivious sorting, which takes $O(n)$ memory accesses
by the algorithm of~\cite{gm-paodor-11})
or at lower levels in the hierarchy (hence, these pointers don't
change by our move into $T_i$).
Moreover, all the nodes of $W$ move as a group.
Thus, any root-to-leaf path in $B$ must be stored in the
tables $T_1$ to $T_L$, plus the queue $Q$ and stash $S$, in a way
that satisfies our induction hypothesis.

The lookup for an element $x$ now requires searching for $O(\log n)$
nodes of $B$ in our hierarchy, which costs an amortized overhead of
$O(\log n)$ time each.
Thus, each lookup costs us an amortized overhead of $O(\log^2 n)$
time.
The obliviousness of this simulation follows from an argument similar
to that given above for the obliviousness for our method that uses
random hash functions.
Therefore, we can perform a stateless oblivious RAM simulation
without using random hash functions with an amortized time overhead 
of $O(\log^2 n)$, assuming a private workspace of size $O(n^\nu)$ for
some constant $\nu>0$.


\section{Performance}
\label{sec:performance}

We have implemented a preliminary prototype of our method for
oblivious RAM simulation based on pseudorandom functions
(Section~\ref{sec:simulator1}) with the goal of estimating the size of
the stash $S$ needed to avoid failures during the rebuild phase.  A
failure can happen when we move elements from table $T_i$ to $T_{i+1}$
and the stash overflows, in which case we need to rebuild
table~$T_{i+1}$.  In this section we present experimental results and
show that for a small constant $s$, a stash of size $s \log n$ is
enough to avoid failures.

Our prototype simulates the dynamic evolution of the hierarchy of hash
tables during the access and rebuild phases, omitting the steps that
maintain obliviousness (e.g. copying the stash to the client's side).
We maintain a stash $S$, a cache $Q$ of size $\log n$ and a hierarchy of $O(\log n)$
cuckoo hash tables $T_1$ to $T_L$, where $L$ is the smallest $i$
such that $2^i \log n \ge n$.
$T_i$ consists of two hash tables of size $(1+\epsilon)2^i \log n$ with hash
functions $h^i_1$ and $h^i_2$. 
Every memory access is followed by the insertion of the corresponding
element into~$Q$. If the item was retrieved from stash $S$,
it is first copied to $Q$ and then is removed from $S$.
We move all the elements of $Q$ to $T_1$ when the number of performed accesses is a multiple
of~$\log n$. 
Similarly, we move all the elements of $T_i$ to $T_{i+1}$ when the number of performed accesses is
a multiple of $2^i \log n$.  New hash functions are
picked for both tables during this phase.  During the insertion, an
item is placed into the stash if after $c \log n$ moves it has not
found an empty cell in the table;  we experiment using $c=2$.
We insert an item into the stash only if it is not already present there.

Our prototype is implemented in Java.
To generate hash functions we use a variation of a method recommended in~\cite{dp-gboehf-08},
where $h^i_1(x) = \mathrm{SHA256}(x~||~\mathrm{seed}^1_i)~\mathrm{mod}~n$, and similarly for $h^i_{2}$.
The seeds are 64-bit long and were obtained using a SHA256 hash chain starting from
an initial seed.

We emphasize that in any implementation of our method there are
various tradeoffs.  For example, increasing the space (that is, using
larger values of $\epsilon$) reduces the average time for an insertion
and the failure probability, as it reduces the frequency with which
items have to be put in the stash.  Increasing the stash size reduces
the failure probability at the expense of additional time to examine
the stash at each step.  Increasing the number of moves allowed before
placing an item in the stash increases the time but lowers the failure
probability.  Our purpose here is not to explore this broad range of
tradeoffs, but to demonstrate the feasibility of this approach;  exploring
finer tradeoffs is left as future work. 

\begin{figure}[htb]
\begin{center}
 \includegraphics[width=60mm,height=85mm,angle=270]{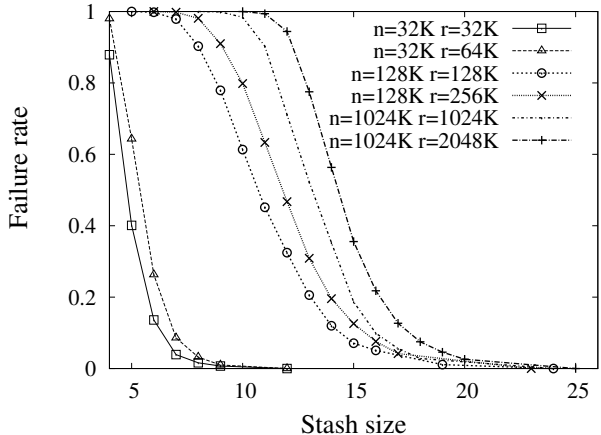}
\caption{Failure rate in 1000 trials for $n$ items and $r$ requests with $c=2$, $\epsilon=0.1$.}
\label{fig:results1}
\end{center}
\end{figure}

\begin{figure}[htb]
\begin{center}
 \includegraphics[width=60mm,height=85mm,angle=270]{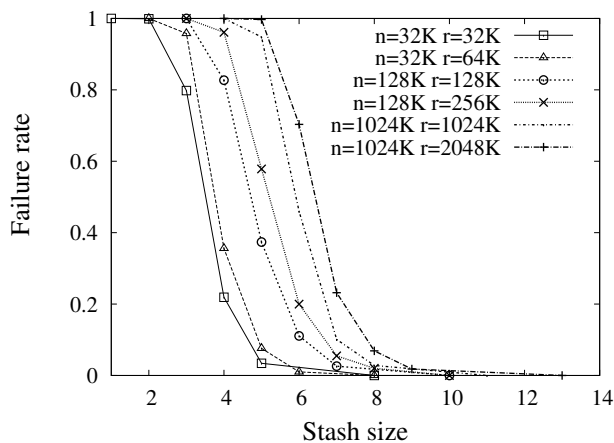}
\caption{Failure rate in 1000 trials for $n$ items and $r$ requests with  $c=2$, $\epsilon=0.2$.}
\label{fig:results2}
\end{center}
\end{figure}

We ran our simulation for up to $1024$K ($\mathrm{K}=1000$) RAM items and a varying number of requests.
Our experiments use a value of $\epsilon$ of 0.1 and 0.2.
For each experiment we recorded the lowest size of $S$ that is needed to avoid a failure.
In Figures~\ref{fig:results1} and~\ref{fig:results2} we show the fraction of trials out of 1000
that result in the stash overflow.
Comparing the two figures, we see that overflows happen substantially more frequently
with $\epsilon = 0.1$ than with $\epsilon = 0.2$, as one would expect since smaller
tables lead to more collisions.
Indeed, for $\epsilon = 0.2$ we found a stash size
of less than $\log n$ was enough to avoid overflows completely in our limited experiments.
Also, a higher number of requests requires a slightly bigger stash since rebuilding happens more
often, leading to a larger maximum stash requirement.  
While much more extensive experimentation would be needed to determine suitable
stash sizes that would avoid stash overflow for numbers of trials many orders of magnitude larger, recall that the probability of an additional item needing to be placed in a stash in a standard cuckoo table falls very quickly.  We thus expect only slightly larger stash sizes for such improvements in robustness.    



\section{Conclusion}
\label{sec:conclusion}
We have given schemes for achieving privacy-preserving access,
with very high probability, to an
outsourced data repository 
for a group of trusted users.
Our scheme assumes each user has a modest amount of private memory,
of size $O(n^\nu)$, for any given fixed constant $\nu>0$,
which is used as a workspace for private computation and carries no state from
one interaction with the data repository to the next.
Assuming the existence of pseudorandom hash functions, say
implemented as keyed SHA-256 functions in practice, our
protocol has an $O(\log n)$ amortized time overhead and an $O(1)$
space overhead.
Moreover, our experiments show that this protocol
would be effective in practice.
If pseudorandom hash functions are not to be used, then we show 
that our protocol can be adapted to have an overhead of $O(\log^2 n)$.

There are several directions for future work, including the
following:
\begin{itemize}
\item
Our protocols assume that the manager of the data 
repository, Bob, is honest-but-curious. 
\begin{itemize}
\item
Can our schemes be
efficiently adapted to the case where Bob is only semi-trusted?
\item
Can we efficiently handle a situation
where Bob acts maliciously against some users?
\item
Can we prevent Bob from performing a replay attack on some users
using old versions of his memory?
\end{itemize}
\item
Our protocols also assume that the members of the group of 
cooperating users are all trusted. 
\begin{itemize}
\item
What if some of the members of the group are
malicious? 
\item
What if some of them collude with Bob to try to reveal the access
patterns of other users?
\end{itemize}
\end{itemize}


\ifAnon\else
\subsection*{Acknowledgments}
This research was supported in part by the National Science
Foundation under grants 0724806, 0713046, 0847968, 0953071, and 1012060.
\fi

{\raggedright
\bibliographystyle{IEEEtranS}
\bibliography{paper}

\begin{thebibliography}{10}
\providecommand{\url}[1]{#1}
\csname url@samestyle\endcsname
\providecommand{\newblock}{\relax}
\providecommand{\bibinfo}[2]{#2}
\providecommand{\BIBentrySTDinterwordspacing}{\spaceskip=0pt\relax}
\providecommand{\BIBentryALTinterwordstretchfactor}{4}
\providecommand{\BIBentryALTinterwordspacing}{\spaceskip=\fontdimen2\font plus
\BIBentryALTinterwordstretchfactor\fontdimen3\font minus
  \fontdimen4\font\relax}
\providecommand{\BIBforeignlanguage}[2]{{%
\expandafter\ifx\csname l@#1\endcsname\relax
\typeout{** WARNING: IEEEtranS.bst: No hyphenation pattern has been}%
\typeout{** loaded for the language `#1'. Using the pattern for}%
\typeout{** the default language instead.}%
\else
\language=\csname l@#1\endcsname
\fi
#2}}
\providecommand{\BIBdecl}{\relax}
\BIBdecl

\bibitem{ahu-dsa-83}
A.~V. Aho, J.~E. Hopcroft, and J.~D. Ullman, \emph{Data Structures and
  Algorithms}.\hskip 1em plus 0.5em minus 0.4em\relax Reading, MA:
  Addison-Wesley, 1983.

\bibitem{aks-osn-83}
M.~Ajtai, J.~Koml{\'o}s, and E.~Szemer{\'e}di, ``An {$O(n \log n)$} sorting
  network,'' in \emph{Proc. ACM Symposium on Theory of Computing (STOC)}, 1983,
  pp. 1--9.

\bibitem{aks-scps-83}
------, ``Sorting in $c\log n$ parallel steps,'' \emph{Combinatorica}, vol.~3,
  pp. 1--19, 1983.

\bibitem{a-orwca-10}
M.~Ajtai, ``Oblivious {RAMs} without cryptographic assumptions,'' in
  \emph{Proc.\ ACM Symposium on Theory of Computing (STOC)}.\hskip 1em plus
  0.5em minus 0.4em\relax ACM, 2010, pp. 181--190.

\bibitem{bm-mrors-10}
P.~Beame and W.~Machmouchi, ``Making {RAMs} oblivious requires superlogarithmic
  overhead,'' Electronic Colloquium on Computational Complexity, Report
  TR10-104, 2010, \url{http://eccc.hpi-web.de/report/2010/104/}.

\bibitem{br-roap-93}
M.~Bellare and P.~Rogaway, ``Random oracles are practical: a paradigm for
  designing efficient protocols,'' in \emph{Proc. ACM Conference on Computer
  and Communications Security (CCS)}.\hskip 1em plus 0.5em minus 0.4em\relax
  New York, NY, USA: ACM, 1993, pp. 62--73.

\bibitem{bmp-rosmor-11}
D.~Boneh, D.~Mazieres, and R.~A. Popa, ``Remote oblivious storage: {M}aking
  oblivious {RAM} practical,'' Technical Report, 2011,
  \url{http://dspace.mit.edu/handle/1721.1/62006}.

\bibitem{cwwz-sclwa-10}
S.~Chen, R.~Wang, X.~Wang, and K.~Zhang, ``Side-channel leaks in web
  applications: a reality today, a challenge tomorrow,'' in \emph{Proc. IEEE
  Symposium on Security and Privacy}, 2010, pp. 191--206.

\bibitem{clrs-ia-01}
T.~H. Cormen, C.~E. Leiserson, R.~L. Rivest, and C.~Stein, \emph{Introduction
  to Algorithms}, 2nd~ed.\hskip 1em plus 0.5em minus 0.4em\relax Cambridge, MA:
  MIT Press, 2001.

\bibitem{dmn-psor-10}
I.~Damg\aa{}rd, S.~Meldgaard, and J.~B. Nielsen, ``Perfectly secure oblivious
  {RAM} without random oracles,'' Cryptology ePrint Archive, Report 2010/108,
  2010, \url{http://eprint.iacr.org/}.

\bibitem{dp-gboehf-08}
Y.~Dodis and P.~Puniya, ``Getting the best out of existing hash functions; or
  what if we are stuck with {SHA}?'' in \emph{Applied Cryptography and Network
  Security (ACNS)}, 2008, pp. 156--173.

\bibitem{gmw-hpamg-87}
O.~Goldreich, S.~Micali, and A.~Wigderson, ``How to play {ANY} mental game,''
  in \emph{Proc. ACM Symposium on Theory of Computing (STOC)}.\hskip 1em plus
  0.5em minus 0.4em\relax New York, NY, USA: ACM, 1987, pp. 218--229.

\bibitem{go-spsor-96}
O.~Goldreich and R.~Ostrovsky, ``Software protection and simulation on
  oblivious {RAMs},'' \emph{J. ACM}, vol.~43, no.~3, pp. 431--473, 1996.

\bibitem{gt-adfai-02}
M.~T. Goodrich and R.~Tamassia, \emph{Algorithm Design: Foundations, Analysis,
  and Internet Examples}.\hskip 1em plus 0.5em minus 0.4em\relax New York, NY:
  John Wiley \& Sons, 2002.

\bibitem{g-rsaso-10}
M.~T. Goodrich, ``Randomized {Shellsort}: {A} simple oblivious sorting
  algorithm,'' in \emph{Proc. ACM-SIAM Symposium on Discrete Algorithms
  (SODA)}.\hskip 1em plus 0.5em minus 0.4em\relax SIAM, 2010, pp. 1--16.

\bibitem{gm-paodor-11}
M.~T. Goodrich and M.~Mitzenmacher, ``Privacy-preserving access of outsourced
  data via oblivious {RAM} simulation,'' in \emph{Proceedings of ICALP}, 2011,
  to appear.

\bibitem{kmw-chs-09}
A.~Kirsch, M.~Mitzenmacher, and U.~Wieder, ``More robust hashing: cuckoo
  hashing with a stash,'' \emph{SIAM J. Comput.}, vol.~39, pp. 1543--1561,
  2009.

\bibitem{kt-ad-05}
J.~Kleinberg and E.~Tardos, \emph{Algorithm Design}.\hskip 1em plus 0.5em minus
  0.4em\relax Boston, MA, USA: Addison-Wesley, Inc., 2005.

\bibitem{k-ss-73}
D.~E. Knuth, \emph{Sorting and Searching}, ser. The Art of Computer
  Programming.\hskip 1em plus 0.5em minus 0.4em\relax Reading, MA:
  Addison-Wesley, 1973, vol.~3.

\bibitem{l-ipaaa-92}
F.~T. Leighton, \emph{Introduction to Parallel Algorithms and Architectures:
  Arrays, Trees, Hypercubes}.\hskip 1em plus 0.5em minus 0.4em\relax San Mateo,
  CA: Morgan-Kaufmann, 1992.

\bibitem{lp-hsn-98}
T.~Leighton and C.~G. Plaxton, ``Hypercubic sorting networks,'' \emph{SIAM J.
  Comput.}, vol.~27, no.~1, pp. 1--47, 1998.

\bibitem{m-soq-09}
M.~Mitzenmacher, ``Some open questions related to cuckoo hashing,'' in
  \emph{Proc. European Symposium on Algorithms (ESA)}, 2009, pp. 1--10.

\bibitem{mu-pcrap-05}
M.~Mitzenmacher and E.~Upfal, \emph{Probability and Computing: Randomized
  Algorithms and Probabilistic Analysis}.\hskip 1em plus 0.5em minus
  0.4em\relax New York, NY, USA: Cambridge University Press, 2005.

\bibitem{pr-ch-04}
R.~Pagh and F.~Rodler, ``Cuckoo hashing,'' \emph{Journal of Algorithms},
  vol.~52, pp. 122--144, 2004.

\bibitem{p-isnod-90}
M.~Paterson, ``Improved sorting networks with {$O(\log N)$} depth,''
  \emph{Algorithmica}, vol.~5, no.~1, pp. 75--92, 1990.

\bibitem{pr-orr-20}
B.~Pinkas and T.~Reinman, ``Oblivious {RAM} revisited,'' in \emph{Advances in
  Cryptology (CRYPTO)}, ser. Lecture Notes in Computer Science, T.~Rabin,
  Ed.\hskip 1em plus 0.5em minus 0.4em\relax Springer, 2010, vol. 6223, pp.
  502--519.

\bibitem{pf-racm-79}
N.~Pippenger and M.~J. Fischer, ``Relations among complexity measures,''
  \emph{J. ACM}, vol.~26, no.~2, pp. 361--381, 1979.

\bibitem{p-ssn-72}
V.~R. Pratt, ``Shellsort and sorting networks,'' Ph.D. dissertation, Stanford
  University, Stanford, CA, USA, 1972.

\bibitem{s-snlgf-09}
J.~Seiferas, ``Sorting networks of logarithmic depth, further simplified,''
  \emph{Algorithmica}, vol.~53, no.~3, pp. 374--384, 2009.

\bibitem{wlgdz-bpstp-10}
G.~Wang, T.~Luo, M.~T. Goodrich, W.~Du, and Z.~Zhu, ``Bureaucratic protocols
  for secure two-party sorting, selection, and permuting,'' in \emph{Proc. ACM
  Symposium on Information, Computer and Communications Security
  (ASIACCS)}.\hskip 1em plus 0.5em minus 0.4em\relax New York, NY, USA: ACM,
  2010, pp. 226--237.

\bibitem{DBLP:conf/ndss/WilliamsS08}
P.~Williams and R.~Sion, ``Usable {PIR},'' in \emph{NDSS}.\hskip 1em plus 0.5em
  minus 0.4em\relax The Internet Society, 2008.

\bibitem{wsc-bcomp-08}
P.~Williams, R.~Sion, and B.~Carbunar, ``Building castles out of mud: practical
  access pattern privacy and correctness on untrusted storage,'' in \emph{Proc.
  ACM Conference on Computer and Communications Security (CCS)}.\hskip 1em plus
  0.5em minus 0.4em\relax New York, NY, USA: ACM, 2008, pp. 139--148.

\end{thebibliography}
}

\end{document}